\documentclass[smallextended]{svjour3}       
\usepackage{amsmath}
\usepackage{amssymb}
\smartqed  
\usepackage{graphicx}

\begin{document}

\title{Discrete Phase Space-Continuous Time Relativistic Klein-Gordon and Dirac Equations, and a New Non-Singular Yukawa Potential}

\author{Anadijiban Das         \and
        Rupak Chatterjee     
}

\institute{Anadijiban Das \at
              Department of Mathematics, Simon Fraser University, Burnaby, British Columbia, V5A 1S6, Canada \\
              \email{Das@sfu.ca}           %  \\
           \and
           Rupak Chatterjee \at
              Center for Quantum Science and Engineering\\
Department of Physics, Stevens Institute of Technology, Castle Point on the Hudson, Hoboken, NJ 07030, USA \\ \email{Rupak.Chatterjee@Stevens.edu}  
}

\date{Received: date / Accepted: date}

\maketitle

\begin{abstract}
This paper deals with the second quantization of interacting \textit{relativistic Fermionic and Bosonic fields} in the arena of discrete phase space and continuous time. The mathematical formulation involves \textit{partial difference equations}. The corresponding Feynman diagrams and a new $S^{\#}$-matrix theory is developed. In the special case of proton-proton M\o ller scattering via an exchange  of a neutral meson, the explicit second order element $\langle f | S^{\#}_{(2)} |i \rangle$ is deduced. In the approximation of very low external three-momenta, \textit{a new Yukawa potential} is explicitly derived from $\langle f | S^{\#}_{(2)} |i \rangle$. Moreover, it is rigorously  proved that this new Yukawa potential is \textit{divergence-free.} The mass parameter of the exchanged meson may be set to zero to obtain a type of scalar Boson exchange between hypothetical Fermions. This provides a limiting case of a new Coulomb type potential directly from the new singularity free Yukawa potential. A divergence-free Coulomb potential between two Fermions at two discrete points is shown to be proportional to the Euler beta function. Within this relativistic discrete phase space continuous time, a single quanta is shown to occupy the hyper-tori $S^{1}_{n^1} \times S^{1}_{n^3} \times S^{1}_{n^3}$ where $S^{1}_{n}$ is a circle of radius $\sqrt{2n+1}$.


\end{abstract}

\section{Introduction}
Quantum mechanics has been exactly represented by the standard partial differential equations of Schr\"{o}dinger, the matrix mechanics of Heisenberg, and the phase space continuous time continuum of Weyl and Wigner \cite{Weyl,Wigner}.  In recent years, an exact representation of quantum mechanics has been introduced in the discrete phase space and continuous time arena \cite{DasI,DasII,DasIII,DasIV,DasV} with the use of partial difference-differential equations.

The second quantization of relativistic interacting fields in a background space-time continuum has given rise to the usual $S$-matrix theory \cite{Jauch,Peskin,Weinberg} and its well known problem of divergences and subsequent renormalization attempts. On the other hand, the analogous $S^{\#}$-matrix in the arena of discrete phase space and continuous time has been shown to soften the the degree of divergences \cite{DasIII,DasIV,DasV,DasVI}. In fact,  in \cite{DasVI}, it was shown the the $S^{\#}$-matrix formulation for a second order electron-electron scattering (or M\o ller scattering) led, in the low momenta approximation of two external electrons, to a new Coulomb potential that was completely divergence-free. 

In this paper, we investigate the problem of a new Yukawa style potential arising from the discrete phase space-continuous time representation of relativistic quantum field theory and the corresponding $S^{\#}$-matrix. We show this new Yukawa style potential is completely  divergence-free. In section 2, we summarize briefly the notations used in the present paper. Section 3 defines various partial difference operators \cite{DasIII,DasIV,DasV,Jordan} and the corresponding basis of Hermite polynomials \cite{Gradshteyn}.
The following three sections describe the second quantization of free relativistic scalar and Fermionic fields and thereafter, interacting fields in the discrete phase space-continuous time $S^{\#}$-matrix formulation \cite{DasIII,DasIV,DasV,Jauch,Peskin,Weinberg,DasVI}. The specific example of M\o ller scattering is investigated.  

Section 7 describes the new discrete phase space Yukawa-style potential. The Green's function of the corresponding partial difference equation in such a static scalar field with Yukawa mass parameter $\mu$ is shown to have a zero discrete phase space distance limit of $\mu \; \exp(\mu^2)\; \Gamma(-1/2, \mu^2)$ where $\Gamma(-1/2, \mu^2)$ is a non-singular incomplete gamma function \cite{Gradshteyn} for $\mu >0$. This divergence free quality is the main attraction of our new Yukawa potential.

In section 8, we take the $\mu=0$ limit of our new Yukawa potential to find a new non-singular Coulomb potential. The Green's function of the corresponding partial difference equation is shown to be proportional to the Euler beta function similar to that found in string theory \cite{Green}. The coincidence discrete space limit is shown to produce a non-singular result similar to our previous work \cite{DasVI}.
We point out that a single field quanta in this relativistic discrete phase space formulation occupy hyper-tori $S^{1}_{n^1} \times S^{1}_{n^3} \times S^{1}_{n^3}$ where $S^{1}_{n}$ is a circle of radius $\sqrt{2n+1}$ as discussed extensively in \cite{DasRCI,DasRCII}. Finally, section 9 summarizes some of the key results of the paper while illustrating the divergence free nature of our new Yukawa style potential.
 
\section{Notations and preliminary definitions}

There exists a characteristic length $l >0$ in this theory, which is conjectured to be the Planck length.  We employ fundamental units characterized by $\hbar = c= l =1$. Thus, all mathematical expressions involving physical phenomena appear as physically dimensionless numbers. Greek indices take values from $\{1,2,3,4\}$ whereas the Roman indices take values from $\{1,2,3\}$. Einstein's summation convention is adopted in both cases. We denote the flat space-time metric by $\eta_{\mu \nu}$ with the corresponding diagonal matrix $[\eta_{\mu \nu}] := diag[1,1,1,-1]$. Therefore, we use a signature of +2 in this paper. An element of our discrete space phase and continuous time is expressed as $(\mathbf{n},x^4) \equiv (n^1, n^2, n^3, t) \in \mathbb N^3 \times \mathbb R, n^j \in \mathbb N$ for $j \in {1,2,3}$ and $x^4 \equiv t \in \mathbb R$.

Let a real or complex-valued function $f$ from $\mathbb N^3 \times \mathbb R$ into $\mathbb R$ or $\mathbb C$ be denoted as $f(\mathbf{n}, t) = f(n^1, n^2, n^3, t)$. We denote the first quantized wave function $\phi(\mathbf{n},t)$ and the second quantized wave function by the same symbol. The context should indicate the quantization order. 

\section{Partial Difference and Differential Operators}

We denote various \textit{partial difference operators and the partial differential operators} as follows \cite{DasIII,DasIV,DasV,DasVI}  

\begin{equation}
\begin{array}{c}
\Delta_j f(\mathbf{n}, t)  := f(...,n^j+1, ..., t) - f(...,n^j, ..., t)   \\ \\
\Delta_j^{'} f(\mathbf{n}, t)  := f(...,n^j, ..., t) - f(...,n^j-1, ..., t)   \\ \\
\Delta_j^{\#} f(\mathbf{n}, t)  := \dfrac{1}{\sqrt{2}} \left[
 \sqrt{n^j+1} f(...,n^j+1, ..., t) 
 - \sqrt{n^j} f(...,n^j-1, ..., t)  \right] \\ \\
 \Delta_j^{o} f(\mathbf{n}, t)  := \dfrac{1}{\sqrt{2}} \left[
 \sqrt{n^j+1} f(...,n^j+1, ..., t) 
 + \sqrt{n^j} f(...,n^j-1, ..., t)  \right] \\ \\
\partial_t f(\mathbf{n}, t)  := \dfrac{\partial}{\partial t} [f(\mathbf{n}, t)]
\end{array}
\end{equation}
We now introduce Hermite polynomials and some pertinent properties by the following equations \cite{DasIII,DasIV,DasV,Gradshteyn},
\begin{equation}
\begin{array}{c}
 H_{n^j}(k_j) := (-1)^{n^j} e^{(k_j)^2}  \dfrac{d^{n^j}}{(dk_j)^{n^j}}[e^{-(k_j)^2}]   \\
\\
\dfrac{d^2}{(dk_j)^2}[H_{n^j}(k_j)] -2k_j\dfrac{d}{dk_j}[H_{n^j}(k_j)]+2n^j [H_{n^j}(k_j)]=0
\\
\\
\dfrac{d}{dk_j}[H_{n^j}(k_j)] = 2n^j [H_{n^j-1}(k_j)], \; n^j \geq 1   \\ \\
H_{n^j+1}(k_j)=2k_j[H_{n^j}(k_j)]-2n^j[H_{n^j-1}(k_j)], \;\;n^j \geq 1
\end{array}
\end{equation}
and the scaled Hermite function
\begin{equation}
\begin{array}{c}
 \xi_{n^j}(k_j)  := \dfrac{(i)^{n^j} e^{-(k_j)^2/2} H_{n^j}(k_j)}{(\pi)^{1/4}2^{(n^j/2)}\sqrt{(n^j)!}} \\ \\
 \displaystyle\int\limits_{{\mathbb R}^3} \left\{ \prod_{j=1}^3 \left[ \xi_{n^j}(k_j)  \overline{\xi_{\hat{n}^j}(k_j)}\right] \right\}   dk_1 dk_2 dk_3  
= \delta_{n^1 \hat{n}^1} \delta_{n^2 \hat{n}^2} \delta_{n^3 \hat{n}^3}  =: \delta^3_{\mathbf{n} \hat{\mathbf{n}}} \\ \\
 -i \Delta^{\#}  \xi_{n^j}(k_j) = k_j \xi_{n^j}(k_j) \\ \\
  -i \Delta^{\#}  \overline{\xi_{n^j}(k_j)} = -k_j \overline{\xi_{n^j}(k_j)} 
\end{array}
\end{equation}

\section{The second quantization of the free relativistic partial difference-differential scalar wave field equation}

The second quantized Klein-Gordon scalar wave field over the discrete phase space-continuous time arena will be denoted as $\phi(\mathbf{n},t)=\phi^{\dagger}(\mathbf{n},t)$. It is Hermitian linear operator acting on a Hilbert space bundle. The linear operator $\phi(\mathbf{n},t))$ satisfies the \textit{partial difference-differential equations} \cite{DasIII,DasIV,DasV}: 
\begin{equation}
\delta^{ab} \Delta_a^{\#}\Delta_b^{ \#}\phi(\mathbf{n},t) -(\partial_t)^2 \phi(\mathbf{n},t) -\mu^2 \phi(\mathbf{n},t) = 0
\end{equation}
where $\mu >0$ denotes the mass of a neutral scalar Boson. A class of exact plane wave solutions of the above equation is furnished by  
\begin{equation}
\begin{array}{c}
\phi^{(-)}(\mathbf{n},t)  = \displaystyle\int\limits_{{\mathbb R}^3} d^3\mathbf{k}\,\,\,[2\omega(\mathbf{k})]^{-1/2} \left\{a(\mathbf{k}) \left[\displaystyle{\prod_{j=1}^{3}} \xi_{n^j}(k_j) \right]  e^{-i\omega t} \right\} \\
\\
\phi^{(+)}(\mathbf{n},t)  = \displaystyle\int\limits_{{\mathbb R}^3} d^3\mathbf{k}\,\,\,[2\omega(\mathbf{k})]^{-1/2} \left\{a^{\dagger}(\mathbf{k}) \left[\displaystyle{\prod_{j=1}^{3}} \overline{\xi_{n^j}(k_j)} \right]  e^{i\omega t} \right\}\\
\\
\phi(\mathbf{n},t)= \phi^{(-)}(\mathbf{n},t)+ \phi^{(+)}(\mathbf{n},t) = \phi^{\dagger}(\mathbf{n},t)\\
\\
\omega \equiv \omega(\mathbf{k}) = k^{4}=-k_{4}=+\sqrt{\mathbf{k}\cdot\mathbf{k}+\mu^2} > 0
\end{array}
\end{equation}
where these improper integrals are supposed to converge uniformly \cite{DasVII,Whittaker}. The linear operator $\phi(\mathbf{n},t)$ representing Bosonic quanta are assumed to possess the relativistic four-momentum $(k^1,k^2,k^3,k^4)=(k_1,k_2,k_3,+\omega)=(\mathbf{k},+\omega)$. 

The canonical quantization rules for operators that act on an infinite dimensional Hilbert space $a_{\mu}(\mathbf{k})$ and $a^{\dagger}_{\mu}(\mathbf{k})$ are assumed to be the commutators:
\begin{equation}
\begin{array}{c}
[a(\mathbf{k}), a^{\dagger}(\hat{\mathbf{k}})]  = -[a^{\dagger}(\hat{\mathbf{k}}), a(\mathbf{k})]= \delta^3 (\mathbf{k}-\hat{\mathbf{k}}) \mathbf{I}(\mathbf{k}) \\
\\
\; [a(\mathbf{k}), a(\hat{\mathbf{k}})]  = [a^{\dagger}(\mathbf{k}), a^{\dagger}((\hat{\mathbf{k}})]=\mathbf{0}(\mathbf{k})
\end{array}
\end{equation} 
where $ \mathbf{I}(\mathbf{k})$ and $\mathbf{0}(\mathbf{k})$ are the identity and zero operators respectively and $\delta^3 (\mathbf{k}-\hat{\mathbf{k}})$ is the Dirac distribution function \cite{DasI,DasII}.
These commutation relations imply the following commutators for our second quantized Bosonic field
\begin{equation}
\begin{array}{c}
[\phi^{(+)}(\mathbf{n},t), \phi^{(+)}(\hat{\mathbf{n}},\hat{t})]  = [\phi^{(-)}(\mathbf{n},t), \phi^{(-)}(\hat{\mathbf{n}},\hat{t})] =\mathbf{0}\\
\\
\;[\phi^{(-)}(\mathbf{n},t), \phi^{(+)}(\hat{\mathbf{n}},\hat{t})]  = -i  \Delta_{(+)}(\mathbf{n},t;\hat{\mathbf{n}},\hat{t};\mu) \mathbf{I} \\
\\
\;[\phi^{(+)}(\mathbf{n},t), \phi^{(-)}(\hat{\mathbf{n}},\hat{t})]  = i \Delta_{(-)}(\mathbf{n},t;\hat{\mathbf{n}},\hat{t};\mu) \mathbf{I} \\
\\
\;[\phi(\mathbf{n},t), \phi^{\dagger}(\hat{\mathbf{n}},\hat{t})]  = -i  \Delta(\mathbf{n},t;\hat{\mathbf{n}},\hat{t};\mu) \mathbf{I} \\
\\
\end{array}
\end{equation}
Here, $\Delta_{(\pm)}(\mathbf{n},t;\hat{\mathbf{n}},\hat{t};\mu)$ and $\Delta(\mathbf{n},t;\hat{\mathbf{n}},\hat{t};\mu)$ are non-singular Green's functions for equation (4) as discussed in the Appendix.  

\section{Second quantization of the free relativistic Fermionic anti-Fermionic partial difference-differential wave equation}

Consider the following irreducible representation of $4 \times 4$ Dirac matrices with real and complex entries \cite{DasVI,DasVII}  

\begin{equation}
\begin{array}{l}
 \gamma^{1}_{(4 \times 4)}:=
\begin{bmatrix}
  0 & 0 & 0 & 1 \\
  0 & 0 & 1 & 0 \\
  0 & 1 & 0 & 0 \\
  1 & 0 & 0 & 0 \\
\end{bmatrix}, \;\;
\gamma^{2}_{(4 \times 4)} :=
\begin{bmatrix}
  0 & 0 & 0 & -i \\
  0 & 0 & i & 0 \\
  0 & -i & 0 & 0 \\
  i & 0 & 0 & 0 \\
\end{bmatrix} \\
\\
 \gamma^{3}_{(4 \times 4)}:=
\begin{bmatrix}
  0 & 0 & 1 & 0 \\
  0 & 0 & 0 & -1 \\
  1 & 0 & 0 & 0 \\
  0 & -1 & 0 & 0 \\
\end{bmatrix}, \;\;
\gamma^{4}_{(4 \times 4)}:= 
\begin{bmatrix}
  -i & 0 & 0 & 0 \\
  0 & -i & 0 & 0 \\
  0 & 0 & i & 0 \\
  0 & 0 & 0 & i \\
\end{bmatrix}
\end{array}
\end{equation}
and their important properties

\begin{equation}
\begin{array}{c}
\gamma^{a\dagger} = \gamma^{a}, \; \gamma^{4\dagger} = -\gamma^{4}\\
\\
\gamma^{\mu} \gamma^{\nu} +  \gamma^{\nu} \gamma^{\mu} = 2 \eta^{\mu \nu} I_{(4 \times 4)}
\end{array}
\end{equation}

The Dirac bispinor wave field $\psi(\mathbf{n},t)=[\psi(\mathbf{n},t)]_{(4 \times 1)}$ is a $4 \times 1$ column vector field with entries of complex numbers. It physically represents a relativistic Fermion-anti-Fermion wave field defined over discrete phase space and continuous time. We introduce a corresponding conjugate   $1 \times 4$ row wave field as follows,
\begin{equation}
\tilde{\psi}(\mathbf{n},t) := i \psi^{\dagger}(\mathbf{n},t) \gamma^{4}
\end{equation}

The discrete phase space continuous time Fermionic-anti-Fermionic wave equations are furnished by \cite{DasVI,DasVII} 
\begin{equation}
\begin{array}{c}
\gamma^a \Delta^{\#}_{a} \psi(\mathbf{n},t) + \gamma^{4} \partial_t \psi(\mathbf{n},t)  + m\psi(\mathbf{n},t) = 0_{(4 \times 1)} \\
\\
\; [\Delta^{\#}_{a} \tilde{\psi}(\mathbf{n},t)]\gamma^{a} + [\partial_t \tilde{\psi}(\mathbf{n},t)] \gamma^{4} -m\tilde{\psi}(\mathbf{n},t) = 0_{(1 \times 4)}
\end{array}
\end{equation}
Here, the positive constant $m>0$ represents the mass of the  Fermionic-anti-Fermionic particle. These equations are essentially the partial difference-differential versions of the Dirac equation.  
We now explore a class of exact plane wave solutions by the following trial solution 
\begin{equation}
\psi(\mathbf{n},t) = \zeta(\mathbf{p}, p_4) \left [\prod_{j=1}^{3} \xi_{n^j}(p_j) \right] e^{ip_4 t} 
\end{equation}
where $\zeta(\mathbf{p}, p_4)$ is a $(4 \times 1)$-column vector function of four-momentum variables $(\mathbf{p}, p_4)$.
By substituting this trial wavefunction into the  partial difference-differential Dirac equation above, we arrive at the following algebraic equations
\begin{equation}
\begin{array}{c}
\eta^{\mu \nu}p_{\mu}p_{\nu} +m^2  = 0,\\
\\
p^4=-p_4=\pm \sqrt{\delta^{ab} p_a p_b +m^2} = \pm \sqrt{||\mathbf{p}||^2 +m^2}\\
\\
E \equiv E(\mathbf{p}) := +\sqrt{\mathbf{p} \cdot \mathbf{p} +m^2} = +\sqrt{||\mathbf{p}||^2 +m^2} >0
\end{array}
\end{equation}
and the following four linearly independent solutions
\begin{equation}
\begin{array}{c}
\zeta_{(r)}(\mathbf{p}, p^4)=\zeta_{(r)}(\mathbf{p}, E)=: u_{(r)}(\mathbf{p}), \;\; E = E(\mathbf{p}) > 0 \\
\\
\zeta_{(r)}(\mathbf{p}, p_4)=\zeta_{(r)}(\mathbf{p}, -E)=: v_{(r)}(\mathbf{p}), \;\; -E = -E(\mathbf{p}) < 0 \\
\\
r \in \{1,2\}
\end{array}
\end{equation} 
where $r \in \{1,2\}$ physically represents the spin-up and down cases for the Fermionic-anti-Fermionic quantas. the four explicit $(4 \times 1)$ column vector solutions are listed below \cite{DasVI,DasVII}
\begin{equation}
\begin{array}{c}
u_{(1)}(\mathbf{p}) = [(m+E)/2E]^{1/2}
\begin{bmatrix}
  1 \\
  0 \\
  -i(m+E)^{-1}p_3  \\
  -i(m+E)^{-1}(p_1 +ip_2)    \\
\end{bmatrix},\\
\\
u_{(2)}(\mathbf{p}) = [(m+E)/2E]^{1/2}
\begin{bmatrix}
  0 \\
  1 \\
  -i(m+E)^{-1}(p_1 -ip_2)  \\
  i(m+E)^{-1}p_3    \\
\end{bmatrix},\\
\\
v_{(1)}(\mathbf{p}) = [(m+E)/2E]^{1/2}
\begin{bmatrix}
  i(m+E)^{-1}p_3  \\
  -i(m+E)^{-1}(p_1 +ip_2)    \\
  1  \\
  0   \\
\end{bmatrix},\\
\\
v_{(2)}(\mathbf{p}) = [(m+E)/2E]^{1/2}
\begin{bmatrix}
  i(m+E)^{-1}(p_1 -ip_2)  \\
  -i(m+E)^{-1}p_3    \\
  0  \\
  1   \\
\end{bmatrix}.
\end{array}
\end{equation} 
Here, $u_{(1)}(\mathbf{p}) $ and $u_{(2)}(\mathbf{p})$ represent Fermionic quanta and $v_{(1)}(\mathbf{p}) $ and $v_{(2)}(\mathbf{p})$ represent anti-Fermionic quanta solutions. 
The above solutions also satisfy the orthonormality conditions
\begin{equation}
\begin{array}{c}
\tilde{u}_{(r)}(\mathbf{p}) \cdot u_{(s)}(\mathbf{p})  
 = - \tilde{v}_{(r)}(\mathbf{p}) \cdot v_{(s)}(\mathbf{p}) =\delta_{(rs)} \\
 \\
\tilde{u}_{(r)}(\mathbf{p}) \cdot v_{(s)}(\mathbf{p})  = \tilde{v}_{(r)}(\mathbf{p}) \cdot u_{(s)}(\mathbf{p}) = 0
\end{array}
\end{equation}

In later sections, we will have to consider the case of very low values of three-momentum $\mathbf{p}$. Therefore, it useful to derive approximations of low external momenta $||\mathbf{p}||$ to the above solutions. The following expansions to order $O \left( ||\mathbf{p}||^4 \right)$  are
\begin{equation}
\begin{array}{c}
 E(\mathbf{p}) = m +\left( ||\mathbf{p}||^2/2m \right) + O \left( ||\mathbf{p}||^4 \right),  \\
 \\
\; [m+E(\mathbf{p})/2E(\mathbf{p})]^{1/2} = [1-(1/2)\left( ||\mathbf{p}||/2m \right)^2 ]+O \left( ||\mathbf{p}||^4 \right)  
 \\ \\
 u_{(1)} = 
\begin{bmatrix}
1\\
0\\
0\\
0
\end{bmatrix} +
\begin{bmatrix}
 -(1/2)\left( ||\mathbf{p}||^2/2m \right)   \\
  0    \\
  -i(p_3/2m)[1-(3/2)\left( ||\mathbf{p}||/2m \right)^2 ] \\
  -i\left(\dfrac{p_1+ip_2}{2m}\right)[1-(3/2)\left( ||\mathbf{p}||/2m \right)^2 ]  \\
\end{bmatrix} + O \left( ||\mathbf{p}||^4 \right) \\
\\
 u_{(2)} =
\begin{bmatrix}
0\\
1\\
0\\
0
\end{bmatrix} +
\begin{bmatrix}
 0  \\
 (1/2)\left( ||\mathbf{p}||^2/2m \right)   \\
  -i\left(\dfrac{p_1-ip_2}{2m}\right)[1-(3/2)\left( ||\mathbf{p}||/2m \right)^2 ] \\
  + i(p_3/2m)[1-(3/2)\left( ||\mathbf{p}||/2m \right)^2 ] \\
\end{bmatrix} + O \left( ||\mathbf{p}||^4 \right)
\end{array}
\end{equation}

A class of exact plane wave solutions of the Dirac partial difference-differential equations above is furnished by the following Fourier-Hermite integrals \cite{DasIII,DasIV,DasV},
\begin{equation}
\begin{array}{c}
\psi_{(4 \times 1)}^{(-)}(\mathbf{n},t) = \displaystyle\int\limits_{{\mathbb R}^3} d^3\mathbf{p}\,\,\,[m/E(\mathbf{p})]^{1/2}  
\left\{ \sum_{r=1}^{2}\alpha_{(r)}(\mathbf{p})u_{(r)}(\mathbf{p}) 
\left(\prod_{j=1}^{3} \xi_{n^j}(p_j) \right)  e^{-iE t} \right\} \\ 
\\
\psi_{(4 \times 1)}^{(+)}(\mathbf{n},t) =  \displaystyle\int\limits_{{\mathbb R}^3} d^3\mathbf{p}\,\,\,[m/E(\mathbf{p})]^{1/2}  
\left\{ \sum_{r=1}^{2}\beta^{\dagger}_{(r)}(\mathbf{p})v_{(r)}(\mathbf{p}) 
\left(\prod_{j=1}^{3} \overline{\xi_{n^j}(p_j)} \right)  e^{iE t} \right\} 
\\  
\\
\psi_{(4 \times 1)}(\mathbf{n},t) =  \psi_{(4 \times 1)}^{(-)}(\mathbf{n},t) + \psi_{(4 \times 1)}^{(+)}(\mathbf{n},t) \\
\\
\tilde{\psi}_{(1 \times 4)}^{(-)}(\mathbf{n},t) = \displaystyle\int\limits_{{\mathbb R}^3} d^3\mathbf{p}\,\,\,[m/E(\mathbf{p})]^{1/2}  
\left\{ \displaystyle\sum_{r=1}^{2}\alpha^{\dagger}_{(r)}(\mathbf{p})\tilde{u}_{(r)}(\mathbf{p}) 
 \left(\prod_{j=1}^{3} \overline{\xi_{n^j}(p_j)} \right)  e^{iE t} \right\} \\ 
\\
\tilde{\psi}_{(1 \times 4)}^{(+)}(\mathbf{n},t) = \displaystyle\int\limits_{{\mathbb R}^3} d^3\mathbf{p}\,\,\,[m/E(\mathbf{p})]^{1/2}  
\left\{\displaystyle\sum_{r=1}^{2}\beta_{(r)}(\mathbf{p})\tilde{v}_{(r)}(\mathbf{p}) 
\left(\prod_{j=1}^{3} \xi_{n^j}(p_j) \right)  e^{-iE t} \right\} 
\\  
\\
\tilde{\psi}_{(1 \times 4)}(\mathbf{n},t) = \tilde{\psi}_{(1 \times 4)}^{(-)}(\mathbf{n},t) +  \tilde{\psi}_{(1 \times 4)}^{(+)}(\mathbf{n},t)
\end{array}
\end{equation}
The 4-component column vector $\psi_{(4 \times 1)}^{(-)}(\mathbf{n},t)$ and the 4-component row vector $\tilde{\psi}_{(1 \times 4)}^{(-)}(\mathbf{n},t)$ are associated with the Fermionic wave field whereas the vectors $\psi_{(4 \times 1)}^{(+)}(\mathbf{n},t)$ and $\tilde{\psi}_{(1 \times 4)}^{(+)}(\mathbf{n},t)$ are the anti-Fermionic wave fields. 

Now, we shall introduce the canonical or second quantization of the free Fermionic-anti-Fermionic wave fields. We adopt a two-dimensional pre-Hilbert space bundle for this purpose. We postulate that the Dirac fields above act as vector bundles in a two-dimensional pre-Hilbert space bundle. Moreover, the Fourier-Hermite coefficients $\alpha_{(r)}(\mathbf{p}), \alpha^{\dagger}_{(r)}(\mathbf{p}), \beta_{(r)}(\mathbf{p}),\beta^{\dagger}_{(s)}(\hat{\mathbf{p}}) $ act as linear operators on this pre-Hilbert space bundle. We assume that these linear operators satisfy anti-commutation rules \cite{Jauch},
\begin{equation}
\begin{array}{c}
\; [A,B]_+  \equiv  AB+BA=[B,A]_+ \\
\\
\; [\alpha_{(r)}(\mathbf{p}), \alpha_{(s)}(\hat{\mathbf{p}})]_+  = [\beta_{(r)}(\mathbf{p}), \beta_{(s)}(\hat{\mathbf{p}})]_+=\mathbf{0} \\
\\
\; [\alpha^{\dagger}_{(r)}(\mathbf{p}), \alpha^{\dagger}_{(s)}(\hat{\mathbf{p}})]_+  = [\beta^{\dagger}_{(r)}(\mathbf{p}), \beta^{\dagger}_{(s)}(\hat{\mathbf{p}})]_+= \mathbf{0}\\
 \\
\; [\alpha_{(r)}(\mathbf{p}), \beta_{(s)}(\hat{\mathbf{p}})]_+  = [\alpha^{\dagger}_{(r)}(\mathbf{p}), \beta^{\dagger}_{(s)}(\hat{\mathbf{p}})]_+= \mathbf{0}\\
\\
\; [\alpha_{(r)}(\mathbf{p}), \beta^{\dagger}_{(s)}(\hat{\mathbf{p}})]_+  = [\alpha^{\dagger}_{(r)}(\mathbf{p}), \beta_{(s)}(\hat{\mathbf{p}})]_+= \mathbf{0}\\
 \\
\; [\alpha_{(r)}(\mathbf{p}), \alpha^{\dagger}_{(s)}(\hat{\mathbf{p}})]_+  = [\beta_{(r)}(\mathbf{p}), \beta^{\dagger}_{(s)}(\hat{\mathbf{p}})]_+ = \delta_{(rs)} \delta^3 (\mathbf{p}-\hat{\mathbf{p}}) \mathbf{I}
\end{array}
\end{equation}

There are key physical motivations for choosing pre-Hilbert space bundle operators $\alpha_{(r)}(\mathbf{p}),\alpha^{\dagger}_{(s)}(\hat{\mathbf{p}}),  \beta_{(r)}(\mathbf{p}), \beta^{\dagger}_{(s)}(\hat{\mathbf{p}})$ satisfying anti-commutation relations in contrast to Hilbert space operators $a_{\mu}(\mathbf{k}), a^{\dagger}_{\nu}(\hat{\mathbf{k}})$ satisfying commutation relations and acting on an infinite dimensional Hilbert space bundle.  These differences are due to the fact that Bosonic fields obey Bose-Einstein statistics, whereas the Fermionic fields obey Fermi-Dirac statistics. 

Now, we present various anti-commutation relations between various quantized Fermionic and anti-Fermionic wave fields,
\begin{equation}
\begin{array}{c}
\;[\psi^{(-)}(\mathbf{n},t), {\psi}^{(+)}(\hat{\mathbf{n}},\hat{t})]_+  = [\psi^{(-)}(\mathbf{n},t), \tilde{\psi}^{(+)}(\hat{\mathbf{n}},\hat{t})]_+ = [0]_{(4 \times 4)} \\
\\
\;[\tilde{\psi}^{(+)}(\mathbf{n},t), {\psi}^{(-)}(\hat{\mathbf{n}},\hat{t})]_+  = [\tilde{\psi}^{(+)}(\mathbf{n},t), \tilde{\psi}^{(-)}(\hat{\mathbf{n}},\hat{t})]_+ = [0]_{(4 \times 4)} \\
\\ \\
\;[\psi^{(-)}(\mathbf{n},t), \tilde{\psi}^{(-)}(\hat{\mathbf{n}},\hat{t})]_+  
= i [S_{(+)}^{\#}(\mathbf{n}, t ;\mathbf{\hat{n}},\hat {t};m)]_{(4 \times 4)} \\
 \\
\;[\tilde{\psi}^{(+)}(\mathbf{n},t), \psi^{(+)}(\hat{\mathbf{n}},\hat{t})]_+  
= i [S_{(-)}^{\#}(\mathbf{n}, t ;\mathbf{\hat{n}},\hat {t};m)]_{(4 \times 4)} \\
 \\
\end{array}
\end{equation}
The various non-singular Green's functions $S_{(a)}^{\#}(\mathbf{n}, t ;\mathbf{\hat{n}},\hat {t};m)$ will be elaborated further in the Appendix.

Let us prove the first non-trivial anti-commutation relation above.
\begin{equation}
\begin{array}{c}
\;[\psi^{(-)}(\mathbf{n},t), \tilde{\psi}^{(-)}(\hat{\mathbf{n}},\hat{t})]_+ =  \displaystyle \sum_{r=1}^{2}\displaystyle \sum_{s=1}^{2} \displaystyle\int\limits_{{\mathbb R}^3} \displaystyle\int\limits_{{\mathbb R}^3} d^3\mathbf{p}d^3\mathbf{\hat{p}}\,\,\,\left[\dfrac{m}{\sqrt{E(\mathbf{p})\hat{E}(\mathbf{\hat{p}})}}\right] \\\\ 
\;[\alpha_{(r)}(\mathbf{p})\alpha^{\dagger}_{(s)}(\mathbf{\hat{p}})]_+[u_{(r)}(\mathbf{p})\tilde{u}_{(s)}(\mathbf{\hat{p}})]  \displaystyle \left(\prod_{j=1}^{3} \xi_{n^j}(p_j)
\overline{\xi_{\hat{n}^j}(\hat{p}_j)} \right) e^{-iEt+i\hat{E}\hat{t}} 
\\\\
or \\\\
\;[\psi^{(-)}(\mathbf{n},t), \tilde{\psi}^{(-)}(\hat{\mathbf{n}},\hat{t})]_+  = \displaystyle \sum_{r=1}^{2}
\displaystyle\int\limits_{{\mathbb R}^3}\left[\dfrac{m}{E(\mathbf{p})}\right][u_{(r)}(\mathbf{p})\tilde{u}_{(r)}(\mathbf{p})] \\ \displaystyle\prod_{j=1}^{3} \xi_{n^j}(p_j) \overline{\xi_{\hat{n}^j}(p_j)}e^{-iE(t-\hat{t})} d^3\mathbf{p}
\end{array}
\end{equation}
Now, it can be shown that \cite{Jauch}
\begin{equation}
\left[\dfrac{m}{E(\mathbf{p})}\right]\displaystyle \sum_{r=1}^{2}[u_{(r)}(\mathbf{p})\tilde{u}_{(r)}(\mathbf{p})] = \left[\dfrac{-i\gamma^j p_j+i\gamma^4E+mI}{2E} \right]_{(4 \times 4)}
\end{equation}
and therefore
\begin{equation}
\begin{array}{c}
\;[\psi^{(-)}(\mathbf{n},t), \tilde{\psi}^{(-)}(\hat{\mathbf{n}},\hat{t})]_+  = \\ \\-
\displaystyle\int\limits_{{\mathbb R}^3} \left[\dfrac{i\gamma^j p_j-i\gamma^4E-mI}{2E} \right] 
\displaystyle \prod_{j=1}^{3} \xi_{n^j}(p_j) \overline{\xi_{\hat{n}^j}(p_j)}e^{-iE(t-\hat{t})} d^3\mathbf{p}
\end{array}
\end{equation}
where in the appendix, it is shown the the Fermionic Green's function is
\begin{equation}
\begin{array}{c}
[S_{(+)}(\mathbf{n}, t ;\mathbf{\hat{n}},\hat {t};m)]_{(4 \times 4)}  = \\ \\i
\displaystyle\int\limits_{{\mathbb R}^3} \left[\dfrac{i\gamma^j p_j-i\gamma^4E-mI}{2E} \right] 
\displaystyle \prod_{j=1}^{3} \xi_{n^j}(p_j) \overline{\xi_{\hat{n}^j}(p_j)}e^{-iE(t-\hat{t})} d^3\mathbf{p}
\end{array}
\end{equation}
Therefore, our result follows.

\section{Interactions of relativistic fields, the $S^{\#}$-matrix in discrete phase space-continuous time, and M\o ller Scattering} 

The relativistic Lagrangian of the second quantized interacting Fermionic and Bosonic fields is assumed to given by \cite{DasIII,DasIV,DasV,Weinberg}
\begin{equation}
L_{int} (\mathbf{n}, t) := -igN [ \tilde{\psi} (\mathbf{n},t)  \psi(\mathbf{n},t) \phi(\mathbf{n},t)].
\end{equation}
Here, $g$ represents the strong coupling constant between the Fermionic-anti-Fermionic field and a Bosonic field. Furthermore, $N[ \cdots ]$ stands for normal ordering. 
The scattering matrix in the background of discrete phase space and continuous time, denoted by $S^{\#}$-matrix, is defined by the operator-valued infinite series \cite{DasIII,DasIV,DasV,Jauch,Peskin,Weinberg}
\begin{equation}
\begin{array}{c}
S^{\#} = I + \sum\limits_{j=1}^{\infty} \dfrac{(g)^j}{j!} \sum\limits_{\mathbf{n}^{1}=(0)} \cdots \sum\limits_{\mathbf{n}^{j}=(0)} \displaystyle \int\limits_{\mathbb{R}} dt^{1}  \\
\\
\cdots \displaystyle\int\limits_{\mathbb{R}} dt^{j}  T \left\{ N[\tilde{\psi} (\mathbf{n}^{1},t_{1}) \psi(\mathbf{n}^{1},t^{1})  \phi (\mathbf{n}^{1},t^{1})]  \right. \\ \\
\cdots \left. N[\tilde{\psi} (\mathbf{n}^{j},t^{j})  \psi(\mathbf{n}^{1},t^{j})  \phi (\mathbf{n}^{j},t^{j})] \right\} =:  I + \displaystyle\sum_{j=1}^{\infty}S^{\#}_{(j)} 
\end{array}
\end{equation}
Here, $T$ denotes Wick's time ordering operation. We distinguish the scattering matrix by the notation $S^{\#}$-matrix from the usual notation of S-matrix in continuous space-time because the physics in the discrete phase space and continuous time is different from the physics in the space-time continuum.

Now, let us consider a particular physical process characterized by the initial Hilbert space state vector $| i \rangle$ and the final Hilbert space state vector $|f \rangle$. The $S^{\#}$-matrix elements for such a process  is provided by 
\begin{equation}
\langle f | S^{\#}| i \rangle = \langle f | i \rangle  + \sum\limits_{j=1}^{\infty}\langle f | S^{\#}_{(j)}| i \rangle 
\end{equation}
Furthermore, we need to define the following vertex distribution functions 
\begin{equation}
\begin{array}{c}
\delta^{\#}_{(3)}(\mathbf{p},\mathbf{q},\mathbf{k}) := \displaystyle \sum_{n^1=0}^{\infty}\sum_{n^2=0}^{\infty}\sum_{n^3=0}^{\infty} \left[\prod_{j=1}^{3} \xi_{n^j}(p_j) \xi_{n^j}(q_j) \xi_{n^j}(k_j) \right] \\ \\
\delta^{\#}_{(3)}(\mathbf{p},-\mathbf{q},-\mathbf{k}) := \displaystyle \sum_{n^1=0}^{\infty}\sum_{n^2=0}^{\infty}\sum_{n^3=0}^{\infty} \left[\prod_{j=1}^{3} \xi_{n^j}(p_j) \overline{\xi_{n^j}(q_j)} \overline{\xi_{n^j}(k_j)} \right]
\end{array} \label{eqn:vertexfunc} 
\end{equation}
The function $\delta^{\#}_{(3)}(\mathbf{p},-\mathbf{q},-\mathbf{k})$ above is different from the standard delta function factor $(2 \pi)^3 \delta^{3}(\mathbf{p},-\mathbf{q},-\mathbf{k})$ found in the usual theory in space-time continuum. Also, the problem of reducing equation (\ref{eqn:vertexfunc}) into much simpler functions is an unsolved problem.

We provide Feynman graph rules to evaluate succinctly each term of the $S^{\#}$-matrix series in Table 1.
\begin{table}
\caption{Feynman Graphs in four-momentum space}
\begin{center}
\includegraphics[scale=0.48]{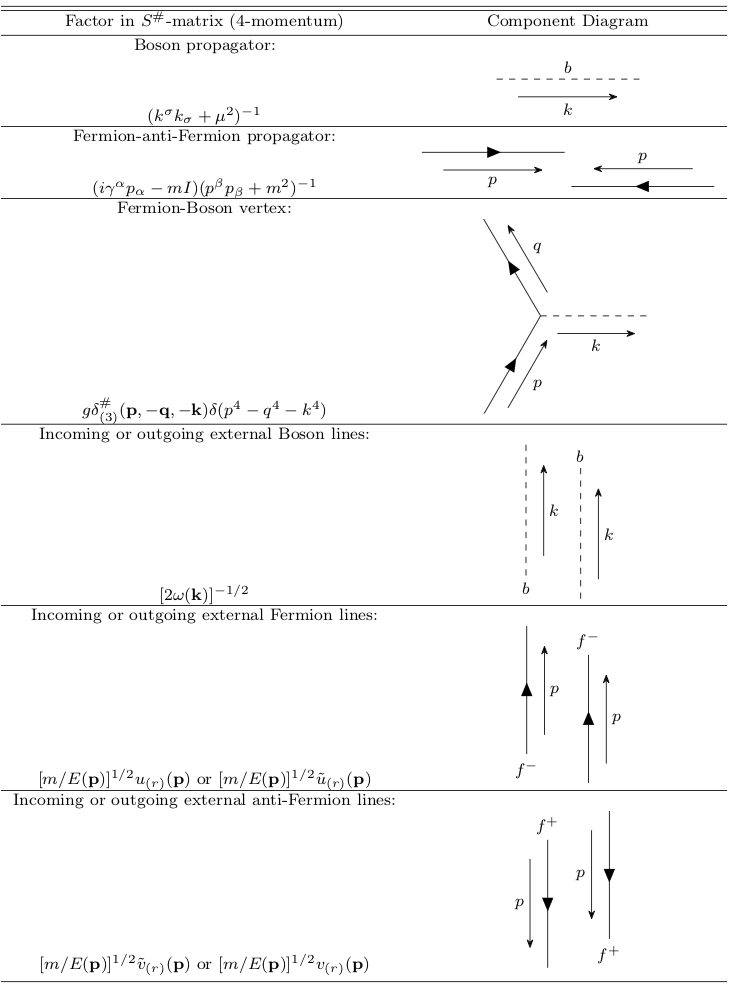}
\end{center}
\end{table}
Note that the relativistic Feynman prescriptions are identical between our $S^{\#}$-matrix and space-time continuum $S$ matrix \textit{except on the vertices}. At a vertex in our formalism, one has the relativistic term
\begin{equation}
g\delta^{\#}_{(3)}(\mathbf{p},-\mathbf{q},-\mathbf{k})\delta(p^4-q^4-k^4).
\end{equation}

As an illustration of computing explicitly an element of$ \langle f | S^{\#}_{(j)} - I | i \rangle$ in (27), using the corresponding component of the Feynman rules in Table 1, we consider the case of M\o ller scattering of two Fermions by exchange of one Boson as depicted in figure 1.
\begin{figure}
\begin{center}
\includegraphics[scale=0.20]{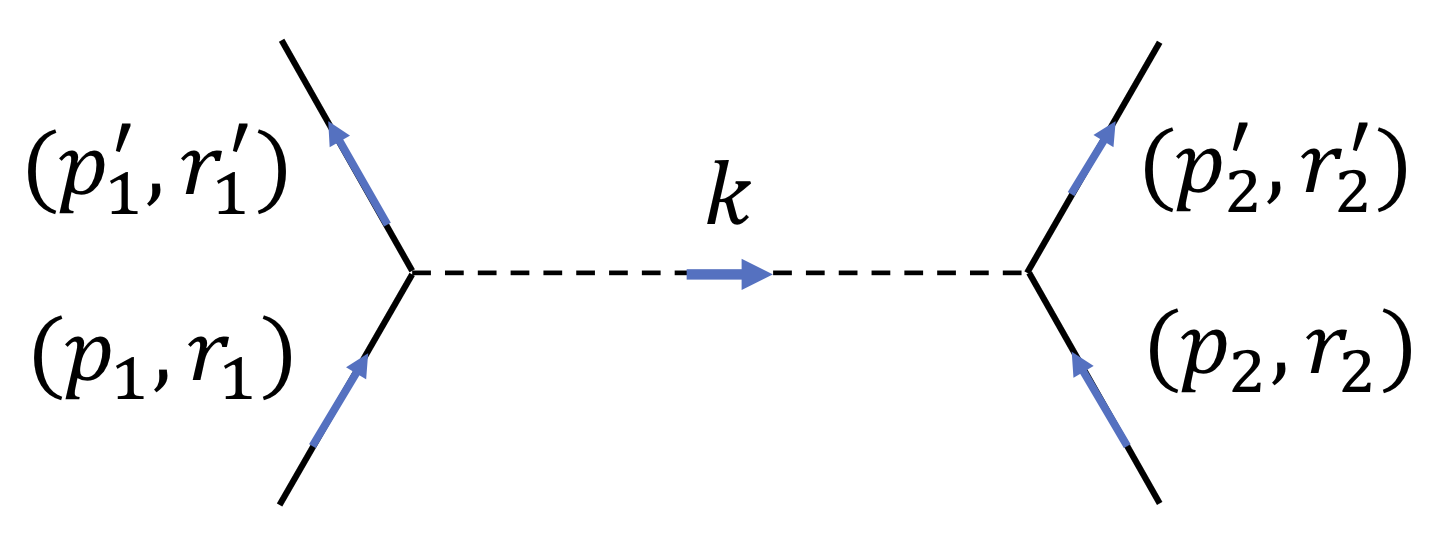}
\end{center}
\caption{The M\o ller scattering of two Fermions by exchange of one Boson}
\end{figure}
The element $\langle p^{'}_2 , p^{'}_1 | S^{\#}_{(2)} | p_1 , p_2  \rangle$ corresponding to figure 1 will be evaluated using the very low momentum approximation for the external Fermion lines as discussed above in (17). Following this, we obtain
\begin{equation}
\begin{array}{c}
\langle p^{'}_2 , p^{'}_1 | S^{\#}_{(2)} | p_1 , p_2  \rangle \approx \left(\dfrac{g^2}{4\pi}\right) \left[ \dfrac{m^2}{\sqrt{E^{'}_1 E^{'}_2 E_1 E_2}} \right] \delta(E^{'}_2+E^{'}_1-E_2-E_1) \\ \\
\;[\delta_{(r'_1,r_1)}\delta_{(r'_2,r_2)}]\displaystyle\int\limits_{{\mathbb R}^3} \left\{ \delta^{\#}_{(3)}(\mathbf{p_1},-\mathbf{p^{'}_1},-\mathbf{k}) [\mathbf{k} \cdot \mathbf{k} +(\mu )^2]^{-1}\delta^{\#}_{(3)}(\mathbf{p_2},-\mathbf{p^{'}_2},+\mathbf{k})\right\} d^3\mathbf{k} \\ \\
= \left(\dfrac{g^2}{4\pi}\right) \left[ \dfrac{m^2}{\sqrt{E^{'}_1 E^{'}_2 E_1 E_2}} \right] \delta(E^{'}_2+E^{'}_1-E_2-E_1)[\delta_{(r'_1,r_1)}\delta_{(r'_2,r_2)}] \\ \\
\displaystyle \int\limits_{{\mathbb R}^3}\biggl\{[\mathbf{k} \cdot \mathbf{k} +\mu^2]^{-1} 
\left[\displaystyle\sum_{\mathbf{n}=0}^{\infty(3)} \prod_{a=1}^{3} \xi_{n^a}(p_{1a}) \overline{\xi_{n^a}(p'_{1a})} \overline{\xi_{n^a}(k_a)} \right] \\ \\
\left[\displaystyle\sum_{\mathbf{\hat{n}}=0}^{\infty(3)} \prod_{b=1}^{3} \xi_{\hat{n}^b}(p_{2b}) \overline{\xi_{\hat{n}^b}(p'_{2b})} \xi_{\hat{n}^b}(k_b)\right]\biggr\}d^3\mathbf{k} 
\end{array}
\end{equation} 
Inside the above equation, one may find  the double Hermite transform for the Green's function defined as
\begin{equation}
G^{\#}(\mathbf{n}, \mathbf{\hat{n}};\mu) := \displaystyle \int\limits_{{\mathbb R}^3}[\mathbf{k} \cdot \mathbf{k} +\mu^2]^{-1} \left[ \displaystyle\prod_{j=1}^{3} \overline{\xi_{n^j}(k_j)}\xi_{\hat{n}^j}(k_j) \right]d^3\mathbf{k}
\end{equation}

To compare and contrast the above result (31) with the usual second order S-matrix elements for M\o ller scattering over space-time continuum, we furnish from \cite{Jauch,DasVI}
\begin{equation}
\begin{array}{c}
\langle p^{'}_2 , p^{'}_1 | S_{(2)} | p_1 , p_2  \rangle \approx \left(\dfrac{g^2}{4\pi}\right) \left[ \dfrac{m^2}{\sqrt{E^{'}_1 E^{'}_2 E_1 E_2}} \right] \delta(E^{'}_2+E^{'}_1-E_2-E_1) \\ \\
\;[\delta_{(r'_1,r_1)}\delta_{(r'_2,r_2)}]\displaystyle\int\limits_{{\mathbb R}^3} \left\{ \delta^3(\mathbf{p_1},-\mathbf{p^{'}_1},-\mathbf{k}) [\mathbf{k} \cdot \mathbf{k} +(\mu )^2]^{-1}\delta^3(\mathbf{p_2},-\mathbf{p^{'}_2},+\mathbf{k})\right\} d^3\mathbf{k} \\ \\
= \left(\dfrac{g^2}{4\pi}\right) \left[ \dfrac{m^2}{\sqrt{E^{'}_1 E^{'}_2 E_1 E_2}} \right] \delta(E^{'}_2+E^{'}_1-E_2-E_1)[\delta_{(r'_1,r_1)}\delta_{(r'_2,r_2)}] \\ \\
\displaystyle \int\limits_{{\mathbb R}^3}\biggl\{[\mathbf{k} \cdot \mathbf{k} +\mu^2]^{-1} 
\left[\dfrac{1}{(2\pi)^{3/2}}\displaystyle \int\limits_{{\mathbb R}^3}\exp\{-i(\mathbf{p_1}-\mathbf{p^{'}_1}-\mathbf{k})\cdot \mathbf{x_1}\} d^3\mathbf{x_1}\right] \\ \\
\left[\dfrac{1}{(2\pi)^{3/2}}\displaystyle \int\limits_{{\mathbb R}^3}\exp\{-i(\mathbf{p_2}-\mathbf{p^{'}_2}-\mathbf{k})\cdot \mathbf{x_2}\} d^3\mathbf{x_2}\right]\biggr\}d^3\mathbf{k} 
\end{array}
\end{equation}
The above equation may be regarded as the double Fourier transform for the Green's function 
\begin{equation}
G(\mathbf{x_1}, \mathbf{x_2};\mu) = G(\mathbf{x_1}-\mathbf{x_2};\mu) := \displaystyle \dfrac{1}{(2\pi)^{3}}\int\limits_{{\mathbb R}^3}[\mathbf{k} \cdot \mathbf{k} +\mu ^2]^{-1} \exp\{-i(\mathbf{x_1}-\mathbf{x_2})\cdot \mathbf{k}\}d^3\mathbf{k}
\end{equation}
with the limit
\begin{equation}
\lim_{\mathbf{x_1} \rightarrow \mathbf{x_2}} |G(\mathbf{x_1}-\mathbf{x_2};\mu) | \rightarrow \infty
\end{equation}
This equation implies the following static partial differential equation
\begin{equation}
\begin{array}{c}
\delta^{ab} \partial_{(1)^a}\partial_{(1)^b}G(\mathbf{x_1}-\mathbf{x_2};\mu)  - \mu^2 G(\mathbf{x_1}-\mathbf{x_2};\mu) \\ \\
= \delta^{ab} \partial_{(2)^a}\partial_{(2)^b}G(\mathbf{x_1}-\mathbf{x_2};\mu)  - \mu^2 G(\mathbf{x_1}-\mathbf{x_2};\mu) = -\delta^{(3)}(\mathbf{x_1}-\mathbf{x_2})
\end{array}
\end{equation}
 
Now, the potential between two Fermions exchanging one Boson is furnished by,
\begin{equation}
V(\mathbf{x_1}-\mathbf{x_2}) = g^2 G(\mathbf{x_1}-\mathbf{x_2};\mu^2) =\dfrac{g^2}{(2\pi)^2} \int\limits_{{\mathbb R}^3}[\mathbf{k} \cdot \mathbf{k} +\mu ^2]^{-1} \exp\{-i(\mathbf{x_1}-\mathbf{x_2})\cdot \mathbf{k}\}d^3\mathbf{k}
\end{equation}
To obtain a closed form solution, we transform to spherical polar coordinates
\begin{equation}
\begin{array}{c}
\mathbf{x}=\mathbf{x_1}-\mathbf{x_2}, \;\;\; r =|| \mathbf{x} ||, \;\;\; k =|| \mathbf{k} || \\\\
\mathbf{x} = (r\sin\theta\cos\phi, r \sin\theta\sin\phi, r \cos\theta) \\\\
\mathbf{k} = (k\sin\hat\theta\cos\hat\phi, k \sin\hat\theta\sin\hat\phi, k \cos\hat\theta) \\\\
\cos \gamma = \sin \theta \sin \hat \theta  \cos \phi \cos \hat \phi +\sin \theta \sin \hat \theta \sin \phi \sin \hat \phi +\cos \theta \cos \hat \theta, \;\;\; 0 \leq \gamma \leq \pi \\\\
\mathbf{x} \cdot \mathbf{k}=rk \cos \gamma =: rky
\end{array}
\end{equation}
thereby obtaining
\begin{equation}
V(r) = \dfrac{g^2}{(2\pi)^2} \int\limits_0^\infty \int\limits_0^\pi \int\limits_{-\pi}^\pi  [\mathbf{k} \cdot \mathbf{k} +\mu ^2]^{-1} e^{irk\cos \gamma}k^2 \sin \gamma dkd\gamma d\phi
\end{equation}
or 
\begin{equation}
\begin{array}{c}
V(r) = \dfrac{g^2}{(2\pi)^2} \displaystyle \int\limits_0^\infty \int\limits\limits_{-1}^1  [\mathbf{k} \cdot \mathbf{k} +\mu ^2]^{-1} e^{irky}k^2 dk dy \\\\
=-i \dfrac{g^2}{(2\pi)^2} \left( \dfrac{1}{r} \right) \displaystyle \int\limits_0^\infty [\mathbf{k} \cdot \mathbf{k} +\mu ^2]^{-1} [e^{irk}-e^{-irk}]k dk \\\\
=  -i \dfrac{g^2}{(2\pi)^2} \left( \dfrac{1}{r} \right) \displaystyle \int\limits_0^\infty [\mathbf{k} \cdot \mathbf{k} +\mu ^2]^{-1} [2i\sin(rk)]k dk\\ \\ 
\end{array}
\end{equation}
Note that $[\mathbf{k} \cdot \mathbf{k} +\mu ^2]^{-1} [2i\sin(rk)]k$ is an even function whereas $[\mathbf{k} \cdot \mathbf{k} +\mu ^2]^{-1} [2i\cos(rk)]k$ is an odd function of $k$. Therefore we may write the above as
\begin{equation}
V(r)=-i \dfrac{g^2}{(2\pi)^2} \left( \dfrac{1}{r} \right) \displaystyle \int\limits_{-\infty}^\infty k[\mathbf{k} \cdot \mathbf{k} +\mu ^2]^{-1} e^{irk} dk 
\end{equation}
The integrand has two simple poles at $k \pm i\mu$.  Using a standard counter-clockwise contour $C$ around the pole $k=i \mu$ in the upper half plane, the integral above using the Cauchy residue theorem is
\begin{equation}
\displaystyle \int\limits_{-\infty}^\infty k[\mathbf{k} \cdot \mathbf{k} +\mu ^2]^{-1} e^{irk} dk  = \oint\limits_{C} k[\mathbf{k} \cdot \mathbf{k} +\mu ^2]^{-1} e^{irk} dk =-i\pi e^{-\mu r}
\end{equation}
Therefore, our potential is
\begin{equation}
V(r)=- \left( \dfrac{g^2}{4\pi}  \right) \dfrac{e^{-\mu r}}{r}
\end{equation}
which is the well know \textit{Yukawa potential} for the strong interaction between two protons exchanging a neutral meson. It is regular for $r >0$ but has a singularity at $r=0$. in Cartesian coordinates, the Yukawa potential for two interacting Fermions is
\begin{equation}
V(\mathbf{x_1},\mathbf{x_2};\mu)=- \left( \dfrac{g^2}{4\pi}  \right) \dfrac{\exp(-\mu||\mathbf{x_1}-\mathbf{x_2}||)}{||\mathbf{x_1}-\mathbf{x_2}||}
\end{equation}
In the case of one Fermion situated at the origin $\mathbf{x_2}=(0,0,0)$ and the other  situated at $\mathbf{x_1}=\mathbf{x}$. the corresponding Yukawa potential is given by,
\begin{equation}
\begin{array}{c}
V(\mathbf{x_1},\mathbf{x_2};\mu)=- \left( \dfrac{g^2}{4\pi}  \right) \dfrac{\exp(-\mu||\mathbf{x}||)}{ ||\mathbf{x}||} \\ \\
=- \left( \dfrac{g^2}{4\pi}  \right) \dfrac{\exp(-\mu\sqrt{(x^1)^2+(x^3)^2+(x^3)^2})}{\sqrt{(x^1)^2+(x^3)^2+(x^3)^2}}
\end{array}
\end{equation}

\section{Discrete phase space and a new non-singular Yukawa potential}

The static approximation to the scalar field equation (4) in the background of discrete phase space is furnished by the following partial difference equation
\begin{equation}
\delta^{ab} \Delta_a^{\#}\Delta_b^{ \#}\phi(\mathbf{n}) -\mu^2 \phi(\mathbf{n}) = 0
\end{equation}
with a Green's function given by
\begin{equation}
\begin{array}{c}
G^{\#}(\mathbf{n}, \mathbf{\hat{n}};\mu) = G^{\#}(n^1,n^2,n^3,\hat{n}^1, \hat{n}^2,\hat{n}^3;\mu) \\ \\
:= \displaystyle \int\limits_{{\mathbb R}^3}[\mathbf{k} \cdot \mathbf{k} +(\mu )^2]^{-1} \left[ \displaystyle\prod_{j=1}^{3} \xi_{n^j}(k_j)\overline{\xi_{\hat{n}^j}(k_j)} \right]d^3\mathbf{k}
\end{array}
\end{equation}
By direct substitution of this Green's function into the partial difference equation (45), one finds, using (2) and (3)
\begin{equation}
\begin{array}{c}
\delta^{ab} \Delta_a^{\#}\Delta_b^{ \#}G^{\#}(\mathbf{n}, \mathbf{\hat{n}};\mu) -\mu^2 G^{\#}(\mathbf{n}, \mathbf{\hat{n}};\mu)  \\ \\
= \delta^{ab} \hat\Delta_a^{\#} \hat\Delta_b^{ \#}G^{\#}(\mathbf{n}, \mathbf{\hat{n}};\mu) -\mu^2 G^{\#}(\mathbf{n}, \mathbf{\hat{n}};\mu)  \\ \\ 
=-\delta_{(n^1 \hat n^1)}\delta_{(n^2 \hat n^2)}\delta_{(n^3 \hat n^3)} =: -\delta^3_{(\mathbf{n} \mathbf{\hat n})}
\end{array}
\end{equation}
Furthermore, from using (2) and (3), one has from (46)
\begin{equation}
\begin{array}{c}
G^{\#}(\mathbf{n}, \mathbf{0};\mu) = \left[ \dfrac{i^{n^1+n^2+n^3}}{\pi^{3/2}2^{(n^1+n^2+n^3)/2}\sqrt{n^1!n^2!n^3!}} \right] \times \\ 
 \displaystyle\int\limits_{{\mathbb R}^3}e^{-\mathbf{k} \cdot \mathbf{k}}[\mathbf{k} \cdot \mathbf{k} +\mu^2]^{-1} \left[ \displaystyle\prod_{j=1}^{3} H_{n^j}(k_j) \right]d^3\mathbf{k}
\end{array}
\end{equation}
and
\begin{equation}
G^{\#}(n^1,0,0,0,0,0;\mu) = \dfrac{i^{n^1}}{\pi^{3/2}2^{\frac{n^1}{2}}\sqrt{n^1!}}  \displaystyle\int\limits_{{\mathbb R}^3}e^{-\mathbf{k} \cdot \mathbf{k}}[\mathbf{k} \cdot \mathbf{k} +\mu^2]^{-1}  H_{n^1}(k_1) d^3\mathbf{k}
\end{equation}

Now, we introduce spherical polar coordinates in three-momentum space
\begin{equation}
\mathbf{k}=(k\cos \theta, k\sin \theta \cos \phi, k\sin \theta \sin \phi), \;\;y=\cos \theta
\end{equation}
to transform (49) into
\begin{equation}
\begin{array}{c}
G^{\#}(n^1,0,0,0,0,0;\mu) = \dfrac{i^{n^1}}{\pi^{3/2}2^{\frac{n^1}{2}}\sqrt{n^1!}} \times \\ \\ \displaystyle\int\limits_{0}^{\infty}\int\limits_{0}^{\pi}\int\limits_{-\pi}^{\pi}  e^{-k^2}
\;[k^2 +\mu^2]^{-1}  H_{n^1}(k\cos \theta)k^2 \sin\theta dk d\theta d\phi \\ \\
=\dfrac{i^{n^1}}{\sqrt{\pi} 2^{\frac{n^1}{2}-1}\sqrt{n^1!}}\displaystyle\int\limits_{0}^{\infty}\int\limits_{-1}^{1}  e^{-k^2}
\;[k^2 +\mu^2]^{-1} H_{n^1}(ky)k^2 dkdy 
\end{array}
\end{equation}
Letting
\begin{equation}
x =k^2 >0, \;\; k=+\sqrt{x} >0
\end{equation} 
(51) becomes
\begin{equation}
G^{\#}(n^1,0,0,0,0,0;\mu) =\dfrac{i^{n^1}}{\sqrt{\pi} 2^{\frac{n^1}{2}}\sqrt{n^1!}}\displaystyle\int\limits_{0}^{\infty}\int\limits_{-1}^{1}  e^{-x}
\;[x +\mu^2]^{-1} H_{n^1}(\sqrt{x}y)\sqrt{x}dxdy 
\end{equation}
The coincidence limit of $n^1 \rightarrow 0$ of (53) yields \cite{Gradshteyn}
\begin{equation}
G^{\#}(0,0,0,0,0,0;\mu) = \mu \; e^{\mu^2}\Gamma\left(-\dfrac{1}{2}, \mu^2 \right)
\end{equation}
Here, the incomplete gamma function is defined by \cite{Gradshteyn}
\begin{equation}
\begin{array}{c}
\Gamma\left(-\dfrac{1}{2}, \mu^2 \right) := \displaystyle\int\limits_{\mu^2}^{\infty} w^{-3/2}e^{-w}dw
\\ \\
\dfrac{\partial}{\partial \mu^2}\Gamma\left(-\dfrac{1}{2}, \mu^2 \right)=-\mu^3 e^{-\mu^2}
\end{array}
\end{equation}
By the above three equations, it has been clearly demonstrated that the function $G^{\#}(0,0,0,0,0,0;\mu)$ is \textit{divergence-free}.

\section{Discrete phase space, a new non-singular Coulomb potential, and Beta functions}

The Coulomb potential may be obtained by setting $\mu=0$ in the previous section. The Green's function from (46) becomes 
\begin{equation}
\begin{array}{c}
G^{\#}(\mathbf{n}, \mathbf{\hat{n}};0) = G^{\#}(n^1,n^2,n^3,\hat{n}^1, \hat{n}^2,\hat{n}^3;0) \\ \\
:= \displaystyle \int\limits_{{\mathbb R}^3}[\mathbf{k} \cdot \mathbf{k}]^{-1} \left[ \displaystyle\prod_{j=1}^{3} \xi_{n^j}(k_j)\overline{\xi_{\hat{n}^j}(k_j)} \right]d^3\mathbf{k}
\end{array}
\end{equation}
or using Hermite polynomials
\begin{equation}
\begin{array}{c}
G^{\#}(\mathbf{n}, \mathbf{0};0) = \left[ \dfrac{i^{n^1+n^2+n^3}}{\pi^{3/2}2^{(n^1+n^2+n^3)/2}\sqrt{n^1!n^2!n^3!}} \right] \times \\ 
 \displaystyle\int\limits_{{\mathbb R}^3}e^{-\mathbf{k} \cdot \mathbf{k}}[\mathbf{k} \cdot \mathbf{k}]^{-1} \left[ \displaystyle\prod_{j=1}^{3} H_{n^j}(k_j) \right]d^3\mathbf{k}
\end{array}
\end{equation}
and
\begin{equation}
G^{\#}(n^1,0,0,0,0,0;0) = \dfrac{i^{n^1}}{\pi^{3/2}2^{\frac{n^1}{2}}\sqrt{n^1!}}  \displaystyle\int\limits_{{\mathbb R}^3}e^{-\mathbf{k} \cdot \mathbf{k}}[\mathbf{k} \cdot \mathbf{k}]^{-1}  H_{n^1}(k_1) d^3\mathbf{k}
\end{equation}
Once again, using the spherical coordinates of (50), we deduce that
\begin{equation}
G^{\#}(n^1,0,0,0,0,0;0) 
=\dfrac{i^{n^1}}{\sqrt{\pi} 2^{(\frac{n^1}{2}-1)}\sqrt{n^1!}}\displaystyle\int\limits_{0}^{\infty}\int\limits_{-1}^{1}  e^{-k^2}
H_{n^1}(ky) dkdy 
\end{equation}
and
\begin{equation}
G^{\#}(2n^1,0,0,0,0,0;0) 
=\dfrac{(-1)^{n^1}}{\sqrt{\pi} 2^{(n^1-1)}\sqrt{(2n^1)!}}\displaystyle\int\limits_{0}^{\infty}\int\limits_{-1}^{1}  e^{-k^2}
H_{2n^1}(ky) dkdy 
\end{equation}
and
\begin{equation}
\begin{array}{c}
G^{\#}(2n^1+1,0,0,0,0,0;0) 
=\dfrac{i(-1)^{n^1}}{\sqrt{\pi} 2^{(n^1-1/2)}\sqrt{(2n^1+1)!}} \times \\
\displaystyle\int\limits_{0}^{\infty}\int\limits_{-1}^{1}  e^{-k^2}
H_{2n^1+1}(ky) dkdy
\end{array}
\end{equation}
As $H_{2n^1}(ky)$ and $H_{2n^1+1}(ky)$ are even and odd functions respectively of the variable $ky$, it is clear that 
\begin{equation}
\begin{array}{c}
\displaystyle\int\limits_{0}^{\infty}\int\limits_{-1}^{1}  e^{-k^2}
H_{2n^1}(ky) dkdy = \dfrac{1}{2}\displaystyle\int\limits_{-\infty}^{\infty}\int\limits_{-1}^{1}  e^{-k^2}H_{2n^1}(ky) dkdy\\ \\ 
and \\ \\
\displaystyle\int\limits_{0}^{\infty}\int\limits_{-1}^{1}  e^{-k^2}
H_{2n^1+1}(ky)(ky) dkdy =0
\end{array}
\end{equation}
Furthermore, from \cite{Gradshteyn}, using the relation
\begin{equation}
\displaystyle\int\limits_{-\infty}^{\infty} e^{-k^2}
H_{2n^1}(ky) dk = (-1)^{n^1}\sqrt{\pi}\dfrac{(2n^1)!}{(n^1!)}(1-y^2)^{n^1}\\ \\ 
\end{equation}
our Green's functions reduce to
\begin{equation}
\begin{array}{c}
G^{\#}(2n^1,0,0,0,0,0;0) 
=\left[ \dfrac{\sqrt{(2n^1)!}}{2^{n^1}(n^1!)}\right]\displaystyle\int\limits_{-1}^{1}(1-y^2)^{n^1}dy \\ \\ 
and \\ \\
G^{\#}(2n^1+1,0,0,0,0,0;0) = 0
\end{array}
\end{equation}
The integral above is the Euler beta function \cite{Gradshteyn}
\begin{equation}
\begin{array}{c}
\displaystyle\int\limits_{-1}^{1}(1-y^2)^{n^1}dy  
= 2^{(2n^1+1)}B(n^1+1,n^1+1)\\\\
=2^{(2n^1+1)} \dfrac{[\Gamma(n^1+1)]^2}{\Gamma(2n^1+2)} 
=2^{(2n^1+1)}\dfrac{[(n^1)!]^2}{[(2n^1+1)!]}
\end{array} 
\end{equation}
leaving us with
\begin{equation}
G^{\#}(2n^1,0,0,0,0,0;0) 
=\dfrac{2^{(n^1+1)}\;(n^1)! \;\sqrt{(2n^1)!}}{(2n^1+1)!}=\dfrac{2^{(n^1+1)}\;(n^1)!}{(2n^1+1)\sqrt{(2n^1)!}} 
\end{equation}
as was initially derived in \cite{DasVI}. In the coincidence limit $n^1 \rightarrow 0_+$, we have 
\begin{equation}
G^{\#}(0,0,0,0,0,0;0) = +2
\end{equation}
indicating that the discrete phase space Coulomb potential has no singularity as was derived in a different manner in  \cite{DasVI}.

\section{A summary of various Discrete phase space non-singular potentials and Green's functions}

Here, we summarize the various discrete phase space and continuous time Green's functions and potentials in a special way. From (46), we have 
\begin{equation}
G^{\#}(\mathbf{n}, \mathbf{\hat{n}};\mu) 
= \displaystyle \int\limits_{{\mathbb R}^3}[\mathbf{k} \cdot \mathbf{k} +(\mu )^2]^{-1} \left[ \displaystyle\prod_{j=1}^{3} \xi_{n^j}(k_j)\overline{\xi_{\hat{n}^j}(k_j)} \right]d^3\mathbf{k}
\end{equation}
and it's various special cases
\begin{equation}
\begin{array}{l}
(i) \; G^{\#}(\mathbf{n}, \mathbf{0};\mu) = \left[ \dfrac{i^{n^1+n^2+n^3}}{\pi^{3/2}2^{(n^1+n^2+n^3)/2}\sqrt{n^1!n^2!n^3!}} \right] \times \\ 
 \displaystyle\int\limits_{{\mathbb R}^3}e^{-\mathbf{k} \cdot \mathbf{k}}[\mathbf{k} \cdot \mathbf{k} +\mu^2]^{-1} \left[ \displaystyle\prod_{j=1}^{3} H_{n^j}(k_j) \right]d^3\mathbf{k} \\ \\
(ii) \; G^{\#}(n^1,0,0,0,0,0;\mu) = \dfrac{i^{n^1}}{\pi^{3/2}2^{\frac{n^1}{2}}\sqrt{n^1!}}  \displaystyle\int\limits_{{\mathbb R}^3}e^{-\mathbf{k} \cdot \mathbf{k}}[\mathbf{k} \cdot \mathbf{k} +\mu^2]^{-1}  H_{n^1}(k_1) d^3\mathbf{k} \\ \\
(iii) \; G^{\#}(2n^1,0,0,0,0,0;\mu) = \dfrac{(-1)^{n^1}}{\pi^{3/2}2^{n^1}\sqrt{(2n^1)!}}  \displaystyle\int\limits_{{\mathbb R}^3}e^{-\mathbf{k} \cdot \mathbf{k}}[\mathbf{k} \cdot \mathbf{k} +\mu^2]^{-1}  H_{2n^1}(k_1) d^3\mathbf{k} \\ \\
(iv) \; G^{\#}(2n^1+1,0,0,0,0,0;\mu)=0 \\ \\
(v) \; G^{\#}(\mathbf{n}, \mathbf{0};0) = \left[ \dfrac{i^{n^1+n^2+n^3}}{\pi^{3/2}2^{(n^1+n^2+n^3)/2}\sqrt{n^1!n^2!n^3!}} \right] \times \\ 
 \displaystyle\int\limits_{{\mathbb R}^3}e^{-\mathbf{k} \cdot \mathbf{k}}[\mathbf{k} \cdot \mathbf{k}]^{-1} \left[ \displaystyle\prod_{j=1}^{3} H_{n^j}(k_j) \right]d^3\mathbf{k} \\ \\
(vi) \; G^{\#}(n^1,0,0,0,0,0;0) = \dfrac{i^{n^1}}{\pi^{3/2}2^{\frac{n^1}{2}}\sqrt{n^1!}}  \displaystyle\int\limits_{{\mathbb R}^3}e^{-\mathbf{k} \cdot \mathbf{k}}[\mathbf{k} \cdot \mathbf{k}]^{-1}  H_{n^1}(k_1) d^3\mathbf{k} \\ \\
(vii) \; G^{\#}(2n^1,0,0,0,0,0;0) = \dfrac{(-1)^{n^1}}{\pi^{3/2}2^{n^1}\sqrt{(2n^1)!}}  \displaystyle\int\limits_{{\mathbb R}^3}e^{-\mathbf{k} \cdot \mathbf{k}}[\mathbf{k} \cdot \mathbf{k}]^{-1}  H_{2n^1}(k_1) d^3\mathbf{k} \\ \\
(viii) \; G^{\#}(2n^1+1,0,0,0,0,0;0)=0
\end{array}
\end{equation}

\begin{figure}
\begin{center}
\includegraphics[scale=0.40]{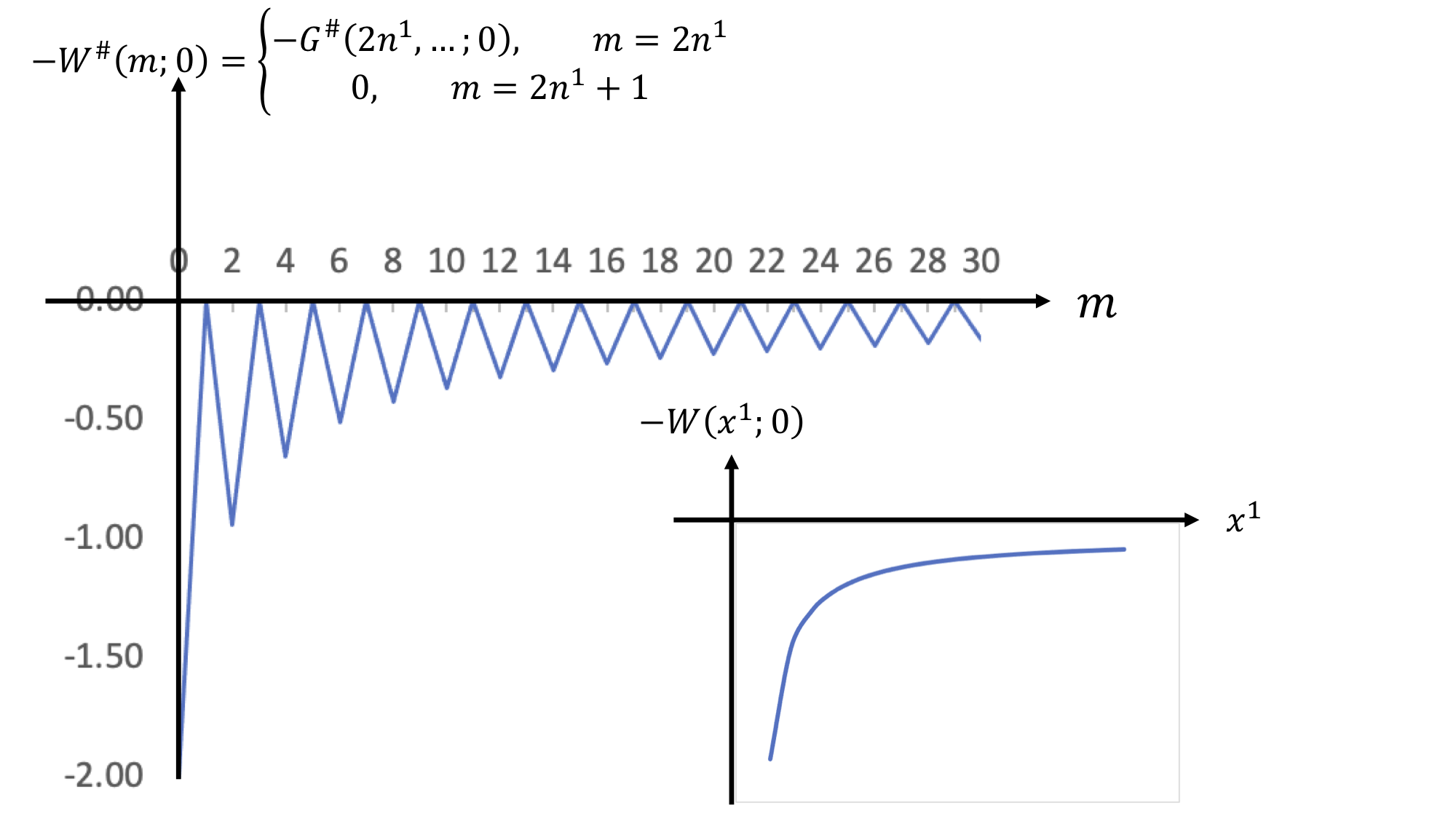}
\end{center}
\caption{The negative of the usual singular Coulomb potential function $W(x^1;0)$ versus its non-singular counterpart $W^{\#}(m;0)$ }
\end{figure}

To reduce the notation, let us define $W^{\#}(n,\mu) := G^{\#}(n,0,0;0,0,0;\mu)$, i.e.
$W^{\#}(2n^1,\mu)=G^{\#}(2n^1,0,0;0,0,0;\mu)$ and compare the discrete formalism to the space-time continuum function implied by equation (42)
\begin{equation}
W(x^1;\mu)=G(x^1,0,0;0,0,0; \mu) = \dfrac{1}{4\pi} \; \dfrac{e^{-\mu |x^1|}}{|x^1|}
\end{equation}
and the Coulomb potential
\begin{equation}
W(x^1;0)=G(x^1,0,0;0,0,0;0) = \dfrac{1}{4\pi} \; \dfrac{1}{|x^1|}
\end{equation}
Note that the actual potential function $V$ between two Fermions interacting via a Boson must include the coupling coefficient $g$ and $e$, i.e.
\begin{equation}
V(x^1;\mu)= W(x^1;\mu)= \dfrac{e^2}{4\pi} \; \dfrac{e^{-\mu |x^1|}}{|x^1|}
\end{equation}
and
\begin{equation}
V^{\#}(n^1;\mu)=-g^2  W^{\#}(n^1,\mu)=  \dfrac{-g^2  \;i^{n^1}}{\pi^{3/2}2^{\frac{n^1}{2}}\sqrt{n^1!}}  \displaystyle\int\limits_{{\mathbb R}^3}e^{-\mathbf{k} \cdot \mathbf{k}}[\mathbf{k} \cdot \mathbf{k} +\mu^2]^{-1}  H_{n^1}(k_1) d^3\mathbf{k}
\end{equation}
It is clear that continuum space-time function $W(x^1;\mu)= \frac{1}{4\pi} \; \frac{e^{-\mu |x^1|}}{|x^1|}$ from above has a singularity at $x^1=0$ whereas the discrete phase space version
\begin{equation}
W^{\#}(2n^1;0) 
=\dfrac{2^{(n^1+1)}\;(n^1)!}{(2n^1+1)\sqrt{(2n^1)!}}
\end{equation} 
has no singularities even at $n^1 \rightarrow 0_+$ as shown in figure 2. We have already shown that $W^{\#}(n^1;\mu)$ is non-singular in equations (54) and (55).

\section*{Appendix: Discrete phase space-continuous time non-singular Green's functions for Fermionic-anti-Fermionic relativistic field equations}

Recall that the relativistic Boson field equation in the discrete phase space-continuous time formalism is given by  
\begin{equation}
\delta^{ab} \Delta_{(a)}^{\#}\Delta_{(b)}^{ \#} \phi(\mathbf{n},t) -(\partial_t)^2 \phi(\mathbf{n},t) -\mu^2 \phi(\mathbf{n},t) = 0
\end{equation}
where $\mu >0$ is the mass of the Boson particle. The associated Green's function are given by \cite{DasIII,DasIV,DasV},
\begin{equation}
\begin{array}{c}
\Delta_{(a)}^{\#}(\mathbf{n},t; \mathbf{\hat{n}},\hat{t};\mu) = (2\pi)^{-1} \displaystyle\int_{{\mathbb R}^3} \left\{
 \left[\prod_{j=1}^{3} \xi_{n^j}(k_j) \overline{\xi_{\hat{n}^j}(k_j)}  \right] \right. \\
\left. \displaystyle\int_{C_{(a)}} \left[ \left( \eta^{\alpha \beta} k_{\alpha} k_{\beta} +\mu^2 \right)^{-1}
\mathbf{exp}(-ik_4(t-\hat{t})]dk^4 \right]  \right\} d^3\mathbf{k} \\ \\
D_{(a)}(\mathbf{n},t;\hat{\mathbf{n}},\hat{t}):=\Delta_{(a)}^{\#}(\mathbf{n},t;\hat{\mathbf{n}},\hat{t}:0) =
\dfrac{1}{2\pi} \displaystyle\int_{{\mathbb R}^3} \left\{
 \left[\prod_{j=1}^{3} \xi_{n^j}(k_j) \overline{\xi_{\hat{n}^j}(k_j)}  \right] \right. \\
\left. \displaystyle\int_{C_{(a)}} \left[ \left( \eta^{\alpha \beta} k_{\alpha} k_{\beta} \right)^{-1}
\mathbf{exp}(-ik_4(t-\hat{t})]dk^4 \right] \right\} d^3\mathbf{k} 
\end{array}
\end{equation}
The Green's functions above involve nine contours in the complex $k^4$-plane as exhibited explicitly in figure 3 with $w = w(\mathbf{k}) := +\sqrt{\mathbf{k} \cdot \mathbf{k} +\mu^2} > 0$.

\begin{figure}
\begin{flushleft}
\includegraphics[scale=0.32]{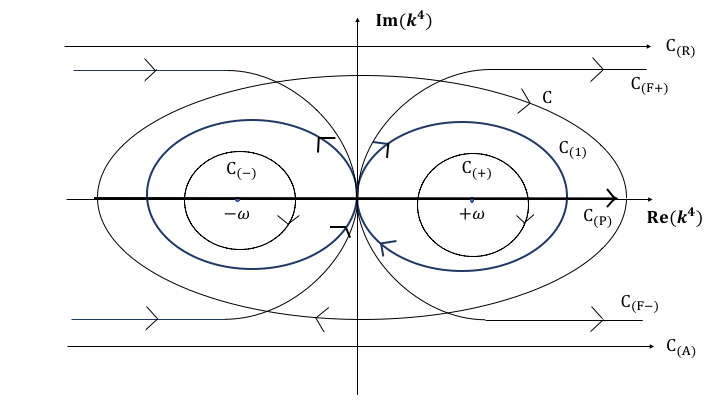}
\end{flushleft}
\caption{Various contours $C_{(a)}$ in the complex $k^4$ plane.}
\end{figure}

\begin{figure}
\begin{flushleft}
\includegraphics[scale=0.32]{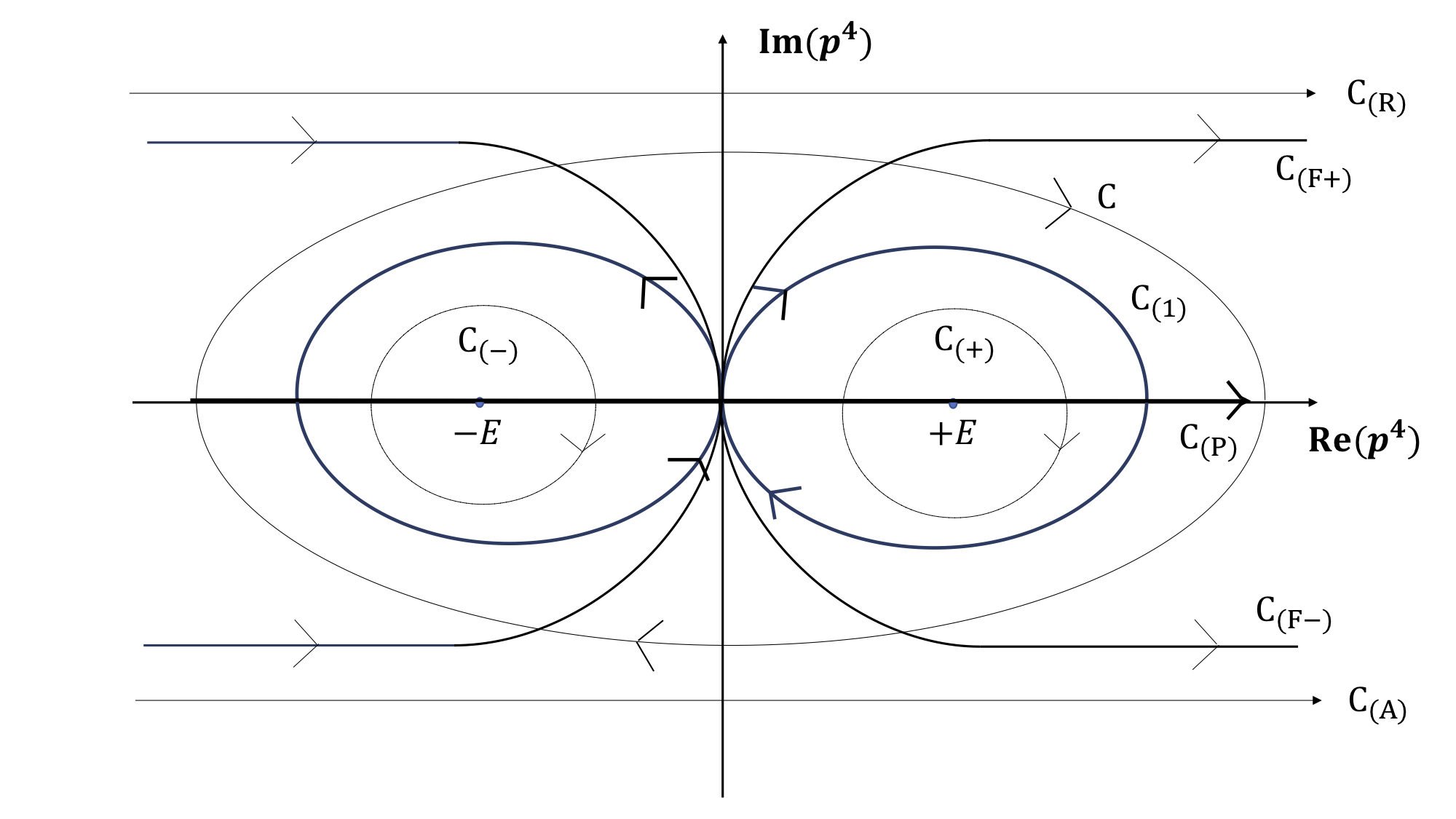}
\end{flushleft}
\caption{Various contours $C_{(a)}$ in the complex $p^4$ plane.}
\end{figure}

Now, recall the Fermion-anti-Fermion wave equations in the discrete phase space-continuous time formalism along with its Green's function is furnished by
\begin{equation}
\begin{array}{c}
\gamma^a \Delta^{\#}_{(a)}\psi(\mathbf{n};t)  + \gamma^{4} \partial_t \psi(\mathbf{n};t) +m\psi(\mathbf{n};t)=[0]_{(4 \times 1)}\\\\
\;[\Delta^{\#}_{(a)}\tilde{\psi}(\mathbf{n};t)]\gamma^a   + [\partial_t \tilde{\psi}(\mathbf{n};t)]\gamma^{4}  - m\psi(\mathbf{n};t)=[0]_{(1 \times 4)} \\ \\
S^{\#}_{(a)}(\mathbf{n},t;\hat{\mathbf{n}},\hat{t};m)_{(4 \times 4)} =(2\pi)^{-1}\displaystyle\int_{{\mathbb R}^3} \left\{ 
\left[\prod_{j=1}^{3} \xi_{n^j}(p_j) \overline{\xi_{\hat{n}^j}(p_j)}  \right] \right. \notag \\
\left. \displaystyle\int_{C{(a)}} \dfrac{(i\gamma^{\mu}p_{\mu} -mI)e^{-ip^4(t-\hat{t})}}{\eta^{\alpha \beta}p_{\alpha}p_{\beta}+m^2}dp^{4} \right\} d^3\mathbf{p}
\end{array}
\end{equation}
The nine $ 4\times 4$ matrix Green's functions above satisfy \cite{DasIII,DasIV,DasV}
\begin{equation}
\begin{array}{c}
(\gamma^j \Delta^{\#}_{j}  + \gamma^{4} \partial_t + mI) S^{\#}_{(a)}(\mathbf{n},t;\hat{\mathbf{n}},\hat{t};m)_{(4 \times 4)} = [0]_{(4\times 4)} \\
\textbf{for contours} \,\,\, C, C_{(\pm)},C_{(1)} \\ \\
 and  \\ \\
= -\delta_{n^1 \hat{n}^1} \delta_{n^2 \hat{n}^2} \delta_{n^3 \hat{n}^3} \delta(t-\hat{t}) [I]_{(4\times 4)}\\
\textbf{for contours} \,\,\, C_{(R)}, C_{(A)},C_{(P)},C_{(F+)},C_{(F-)}  
\end{array}
\end{equation}  
with the contours depicted in figure 4.

\end{document}